\documentclass[letterpaper,twocolumn,english,prb,aps,superbib,tightenlines,floatfix]{revtex4-1}
\usepackage{hyperref,natbib}
\usepackage{graphicx,epsfig}

\begin{document}

\newcommand{\half}{\frac{1}{2}}
\title{Reading, writing and squeezing the entangled states of two nanomechanical resonators coupled to a SQUID}
\author{Guy Z. Cohen}
\email{gcohen@physics.ucsd.edu}
\affiliation{Department of
Physics, University of California, San Diego, La Jolla, California
92093-0319}

\author{Massimiliano Di Ventra}
\email{diventra@physics.ucsd.edu}
\affiliation{Department of
Physics, University of California, San Diego, La Jolla, California
92093-0319}

\begin{abstract}
We study a system of two nanomechanical resonators embedded in a
dc SQUID. We show that the inductively-coupled resonators can be
treated as two entangled quantum memory elements with states that
can be read from, or written on by employing the SQUID as a
displacement detector or switching additional external magnetic
fields, respectively. We present a scheme to squeeze the even mode
of the state of the resonators and consequently reduce the noise
in the measurement of the magnetic flux threading the SQUID. We
finally analyze the effect of dissipation on the squeezing using
the quantum master equation, and show the qualitatively different
behavior for the weak and strong damping regimes. Our predictions
can be tested using current experimental capabilities.
\end{abstract}

\maketitle

\section{Introduction} \label{introduction}
In recent years nanoelectromechanical
systems\cite{Blencowe05,Schwab05} (NEMs), nanoscale mechanical
oscillators coupled to electronic devices of comparable
dimensions, have attracted substantial research effort. A major
motivation for this effort is the ability to observe quantum
behavior in a macroscopic system under realizable experimental
conditions\cite{Leggett86,Leggett02}. Indeed, NEMs today can be
fabricated with vibrational mode frequencies of 1 MHz-10 GHz and
quality factors in the range of $10^3-10^5$, allowing the quantum
regime to be reached at milli-Kelvin temperatures for high
frequency oscillators\cite{LaHaye04,OConnell10}. Possible quantum
effects in NEMs under such conditions include quantized energy
levels, superposition of states, entanglement and
squeezing\cite{Rabl04,Ruskov05,Zhou06}. In addition, NEMs are
applied to high-sensitive detection of
mass\cite{Li04,Buks06-2,Jensen08}, force\cite{Blencowe07} and
displacement\cite{Stampfer06}, electrometers\cite{Blick00}, and
also to classical memory elements\cite{Blick07,Mahboob08}.

Observing or changing the state of NEMs requires some type of
transducer which couples to them. Optical
coupling\cite{Schliesser09} can be performed, e.g., by a microwave
cavity\cite{Regal08}, but is difficult to integrate in circuits
and suffers from the diffraction limit and heating of the NEMS.
Non-optical coupling methods are therefore more common in
experiments today. With magnetomotive coupling\cite{Huang05}, the
magnetic force on a thin metallic layer on the NEMS is measured.
Capacitive coupling can take many forms, one of which uses a
normal or superconducting single electron transistor
(SET)\cite{Mozyrsky04,Ruskov05}. The NEMS changes the island
charging energy in the SET and hence the tunneling rates, which
can be read electronically. Other forms of capacitive coupling use
Cooper pair boxes\cite{Irish03,Suh10}, flux qubits\cite{Xue07-2},
quantum point contacts\cite{Benatov12} and quantum
dots\cite{Lambert08}.

An inductive coupling scheme with a potential for displacement
precision greater than the standard quantum limit is obtained by
integrating a doubly clamped micron-scale beam within a
superconducting quantum interference device (SQUID). In a dc SQUID
the motion of the resonator changes the area of the SQUID loop and
hence the magnetic flux and the current through it, which is then
measured. This system was only recently
implemented\cite{Poot08,Poot10}. A more sophisticated design,
where the dc SQUID, and hence the resonator, is coupled to a
charge qubit, was also proposed\cite{Zhou06}. For an rf SQUID it
was found\cite{Buks06} that the change in the magnetic flux due to
the motion of the beam affects the visibility of Rabi oscillations
in the SQUID levels. The detection of discrete Fock states in a
resonator integrated with an rf SQUID was suggested in another
work\cite{Buks08}.

Squeezed states, originally introduced in quantum
optics\cite{Walls83}, are defined as minimum-uncertainty states
with less noise in one field quadrature than a coherent
state\cite{Walls08}. Several methods to generate squeezing in NEMs
were suggested. Coupling to a charge qubit\cite{Rabl04,Zhou06} as
means of generating squeezing was proposed, while another work
described squeezing by periodic position measurement with a
weakly-coupled detector\cite{Ruskov05}. Squeezing in nanoresonators can
be applied to decrease the noise in force or displacement
measurements to below the standard quantum limit, greatly
improving the sensitivity of the device\cite{Rabl04,Teufel09}.

In this work, we present a scheme to create quantum entanglement
and squeezing in two nanoresonators integrated in a dc SQUID. A
previous study\cite{Xue07} analyzed a similar system but
introduced many approximations that are difficult to implement
experimentally, whereas our present study is closer to an
experimentally realizable system. For instance, we do not overlook
the generally non-negligible self-inductance of the SQUID as done
in previous work\cite{Xue07}, and we assume mega-Hertz frequency
nanomechanical oscillators rather than giga-Hertz frequency
resonators, which are difficult to integrate with a SQUID. Lastly,
we do not require the SQUID to be prepared in a high-$|\alpha|$
coherent state in order to have squeezing, as the previous study
does\cite{Xue07}, and require instead a thermal equilibrium state,
which is easier to accomplish. Finally, we consider different
aspects of the system and draw conclusions, e.g., on the reading
and writing processes, that were not advanced in previous
literature.

The paper is organized as follows. In Sec. \ref{section1} we
present the system model and its classical Lagrangian and
Hamiltonian formulations. We then proceed to quantize the
Hamiltonian for the non-dissipative case and derive the effective
Hamiltonian. Next, in Sec. \ref{section2} we treat the system as a
quantum memory and explain how one can read its quantum state or
write on it. In Sec. \ref{section3} we put forward a scheme for
generating quadrature-squeezed states of the nanomechanical beams
when dissipation is neglected, whereas in Sec. \ref{section4} we
use a quantum master equation approach to test the range of
validity of our results in the the presence of dissipation.
Lastly, we discuss our results and present the conclusions in Sec.
\ref{conclusions}.
\section{System Model and Hamiltonian} \label{section1}

\begin{figure}[t]
 \begin{center}
 \includegraphics[width=8.50cm]{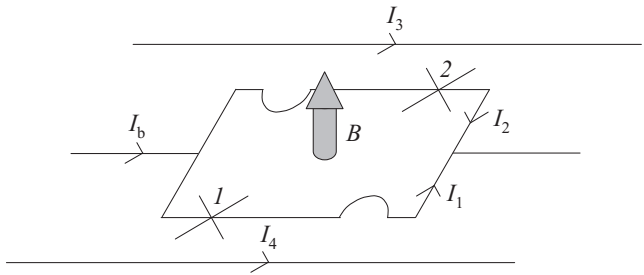}
 \caption{\label{fig01} Schematic of the device we study: Two nanomechanical beams oscillating in plane are embedded in a dc SQUID with area $A$ when beams are at rest. Uniform magnetic field $B$ threads the SQUID, and bias current $I_b$ is assumed. The two identical Josephson junctions on the SQUID have phase drops $\gamma_i$ ($i=1,2$). The currents $I_3$ and $I_4$ create additional magnetic fields $B_1$ and $B_2$ at the resonators.}
 \end{center}
\end{figure}

Our system consists of a dc SQUID, shown schematically in Fig.
\ref{fig01}, in which each arm includes a Josephson junction and
an integrated doubly clamped beam of length $l_i$ and mass $m_i$
that can oscillate mechanically in the plane of the SQUID with an
angular frequency $\widetilde{\omega}_i$ ($i=1,2$). The notation
we use is similar to the one in Ref. \onlinecite{Pugnetti10}. A
uniform magnetic field $B$ is applied perpendicularly to the plane
of the loop, and a dc bias current $I_b$ flows through it after
splitting to $I_1$ and $I_2$ in the lower and upper arms,
respectively. Two current-carrying wires with currents $I_3$ and
$I_4$ create additional magnetic fields $B_1$ and $B_2$ at the
positions of the first and second beams, respectively. The beam
amplitudes are much smaller than the beam-wire distance, allowing
these fields to be approximated as being spatially uniform. For
simplicity, the two Josephson junctions on the SQUID arms are
taken to be identical and their gauge-invariant phase changes are
denoted by $\gamma_i$. The critical current and shunting
capacitance of each junction are taken as $I_c$ and $C$,
respectively, and are used to define the characteristic junction
energy scales: the Josephson energy $E_J=\hbar I_c/2e$ and the
charging energy $E_C=e^2/2C$. The plasma frequency
$\omega_{pl}=\sqrt{2E_JE_C}/\hbar$ sets the typical time scale for
the SQUID dynamics.

The area of the SQUID loop depends on the center of mass positions
of the nanomechanical resonators, denoted by $x_i$ and defined as
zero when the beam is at rest, and positive when it is inside the
loop. Since the superconducting order parameter is single valued,
we must have
\begin{eqnarray}
&\gamma_1-\gamma_2-\frac{2\pi\Phi}{\Phi_0}=2\pi p, \label{eq01-1}\\
&\Phi=BA-\sum\limits_i(B+B_i)l_ix_i+L(I_1-I_2)/2, \label{eq01-2}
\end{eqnarray}
where $p$ is an integer, $\Phi$ is the total magnetic flux
threading the loop, $A$ is the loop area when the beams are at
rest, $\Phi_0=h/2e$ is the flux quantum, $L$ is the
self-inductance of the loop, and $I_i$ is the current in its $i$th
arm. The first term in Eq. (\ref{eq01-2}) comes from the external
magnetic field and the second, responsible for the coupling of the
mechanical and magnetic degrees of freedom, from the oscillation
of the beams. The difference between $l_ix_i$ and the actual area
enclosed by the $i$th beam is negligible, being of third order in
the ratio of the beam amplitude to its length. Lastly, the third
term originates from the magnetic flux induced by the circulating
current in the SQUID.

The kinetic and potential energies of the system are functions of
four dimensionless variables defined by
$\gamma=(\gamma_1+\gamma_2)/2$, $\phi=\Phi/\Phi_0$ and
$\xi_i=(B+B_i)l_ix_i/\Phi_0$. They are
\begin{eqnarray}
&T=\sum\limits_i(\frac{\hbar^2}{4E_C}\frac{1}{\Omega_i^2}\frac{1}{\mathcal{A}_i^2}\dot{\xi}_i^2)+\frac{\hbar^2}{2E_C}\dot{\gamma}^2+\frac{\pi^2\hbar^2}{2E_C}\dot{\phi}^2, \label{eq02-1}\\
&U=E_{J} [-2\cos\gamma \cos(\pi\phi) -
\frac{I_b}{I_c}\gamma+\sum\limits_i ((-1)^i\pi \frac{I_b}{I_c}\xi_i + \frac{\xi_i^2}{2\mathcal{A}_i^2})\nonumber \\
&+\frac{2\pi}{\beta_L}(\phi-\xi_1-\xi_2-\phi_e)^2], \label{eq02-2}
\end{eqnarray}
where mechanical dissipation was assumed to be negligible, and
where we define the screening parameter $\beta_L=2LI_c/\Phi_0$ and
external flux $\phi_e=BA$ for the SQUID, while the dimensionless
magnetic field
\begin{equation}
\mathcal{A}_i=\sqrt{\frac{E_J}{m_i}}\frac{(B+B_i)l_i}{\widetilde{\omega}_i\Phi_0}
\label{eq03}
\end{equation}
and oscillation frequencies
$\Omega_i=\widetilde{\omega}_i/\omega_{pl}$ are defined for each
of the beams. The first term in Eq. (\ref{eq02-1}) corresponds to
the kinetic energy of the beams, while the second and third terms
to the capacitive energy of the junctions. The first term in Eq.
(\ref{eq02-2}) relates to the Josephson junctions energy, while
the second term is the washboard potential term\cite{Poole07}. The
third term corresponds to the Lorentz force on the beams in the
classical equations of motion (EOMs), and the fourth term to the
beams' elastic potential, taken to be harmonic, as nonlinear terms
are negligible at the amplitudes concerned\cite{Aldridge05}.
Lastly, the fifth term corresponds to the inductive energy of the
SQUID.

The classical EOMs for the four variables $\gamma,\phi,\xi_1$ and
$\xi_2$ are the Euler-Lagrange equations for the system Lagrangian
$\mathcal{L}=T-U$. Before writing the Hamiltonian, we expand the
potential in series about a minimum
$(\overline{\phi},\overline{\gamma},\overline{\xi}_1,\overline{\xi}_2)$,
around which the system oscillates. Under current experimental
conditions\cite{Poot08,Poot10} such a minimum exists, as the
Hessian matrix for $U$ there, proportional to the one in Eq.
\ref{eq06}, is positive definite. The well containing the minimum
can accommodate $\sim 20$ states in $\gamma$ and $\sim 900$ in
$\phi$. If we take these two parameters to be ``frozen'' at their
respective ground states, as we will later assume, we find the
well to be infinitely deep for the $\xi_i$ parameters. This
assumption also allows us to neglect in the series expansion of
$U$ terms higher than quadratic ones in $\phi-\overline{\phi}$ and
$\gamma-\overline{\gamma}$. With these approximations in mind the
Hamiltonian $H$ reads
\begin{equation}
H=T+U=\sum\limits_i
\frac{E_C}{2\hbar^2}p_i^2+\sum\limits_{i,j}E_JV_{ij}q_iq_j,\label{eq04}
\end{equation}
where the coordinates $q_i$ are given by
\begin{equation}
q_1=\gamma-\overline{\gamma},\ \ q_2=\pi(\phi-\overline{\phi}),\ \
\
q_{2+j}=\frac{1}{\sqrt{2}\Omega_j\mathcal{A}_j}(\xi_j-\overline{\xi}_j)\label{eq05}
\end{equation}
($j=1,2$), and the canonically conjugate momenta $p_i$ are
$p_i=(\hbar^2/E_C)\dot{q}_i$. In addition,
\begin{equation}
V= \left(
\begin{array}{cccc}
    r & - s & 0 & 0 \\
    - s  & r+\frac{2}{\pi\beta _L} & -\frac{2\sqrt{2}\Omega_1\mathcal{A}_1}{\beta _L} & -\frac{2\sqrt{2}\Omega_2\mathcal{A}_2}{\beta _L} \\
 0 & -\frac{2\sqrt{2}\Omega_1\mathcal{A}_1}{\beta _L} & \Omega_1^2(1+\frac{4 \pi \mathcal{A}_1^2}{\beta _L}) & \frac{4 \pi  \Omega_1\mathcal{A}_1 \Omega_2\mathcal{A}_2}{\beta _L} \\
 0 & -\frac{2\sqrt{2}\Omega_2\mathcal{A}_2}{\beta _L} & \frac{4 \pi  \Omega_1\mathcal{A}_1 \Omega_2\mathcal{A}_2}{\beta _L} & \Omega_2^2(1+\frac{4 \pi \mathcal{A}_2^2}{\beta _L})
\end{array}
\right),\label{eq06}
\end{equation}
where $r=\cos \overline{\gamma} \cos \left(\pi
\overline{\phi}\right)$ and $s=\sin \overline{\gamma} \sin
\left(\pi  \overline{\phi}\right)$ were introduced. We see that
the beam oscillations are coupled inductively via the $V_{34}$
term. This coupling can be used to generate squeezed states in the
beams as we will show below.

The Hamiltonian is quantized in the standard way by converting the
coordinates $q_i$ and their canonically conjugate momenta $p_i$ to
operators and postulating the canonical commutation relation
$[\widehat{q}_i,\widehat{p}_j]=i\hbar\delta_{ij}$. In terms of
creation and annihilation operators $\widehat{q}_i$ and
$\widehat{p}_i$ are
\begin{eqnarray}
&\hat{q}_i=\half(\frac{2E_C}{E_JV_{ii}})^{1/4}(a_i^\dagger+a_i), \label{eq07-1}\\
&\hat{p}_i=i\hbar(\frac{E_JV_{ii}}{2E_C})^{1/4}(a_i^\dagger-a_i),
\label{eq07-2}
\end{eqnarray}
and the quantized Hamiltonian is given by
\begin{eqnarray}
&H=\sum\limits_{i} \hbar\omega_i(a_i^\dagger
a_i+\half)\nonumber\\
&+\frac{1}{4}\hbar\omega_{pl}\sum\limits_{i\neq j}
\frac{\omega_{pl}}{\sqrt{\omega_i\omega_j}}V_{ij}(a_i+a_i^\dagger)(a_j+a_j^\dagger),\label{eq08}
\end{eqnarray}
where $\omega_i=\omega_{pl}\sqrt{V_{ii}}$. We note the frequencies
$\omega_3$ and $\omega_4$ are the same as the resonators
frequencies, $\widetilde{\omega}_1$ and $\widetilde{\omega}_2$,
apart from each having a factor due to the magnetic field at the
resonator. Thus we see that the Lagrangian classical memory
variables $\xi_1$ and $\xi_2$, in complete analogy with memory
variables in electronic circuits\cite{Cohen12}, become memory
quanta in the Hamiltonian.

Taking the same experimental conditions\cite{Poot08,Poot10} and
tuning $B$ to make $r$ of order unity results in $\hbar\omega_1\gg
k_BT$ and $\hbar\omega_2\gg k_BT$. Consequently, the first and
second harmonic oscillators are ``frozen'' at their respective
ground states. Moreover, since $\omega_1,\omega_2\gg
\omega_3,\omega_4$, exciting the nanomechanical oscillators will
not budge them from their ground state. Removing constant terms,
we are then left with the effective Hamiltonian
\begin{equation}
H=\hbar\omega_3 a_3^\dagger a_3+\hbar\omega_4 a_4^\dagger
a_4+\widetilde{V}(a_3+a_3^\dagger)(a_4+a_4^\dagger),\label{eq09}
\end{equation}
where the interaction coefficient reads
\begin{equation}
\widetilde{V}=\frac{2\pi\hbar\sqrt{\widetilde{\omega}_1\widetilde{\omega}_2}\mathcal{A}_1\mathcal{A}_2}{\beta_L(1+\frac{4\pi
\mathcal{A}_1^2}{\beta_L})^{1/4}(1+\frac{4\pi
\mathcal{A}_2^2}{\beta_L})^{1/4}}.\label{eq10}
\end{equation}
\section{Reading and Writing Quantum Information}\label{section2}
We now wish to employ this system to create entangled
nanomechanical quantum memory that can be read from and written
on. We assume the beams are cooled to a temperature low enough so
as to reduce the equilibrium state to the ground state for each of
the beams. This is possible today, e.g. by coupling to a
superconducting microwave resonator\cite{Teufel08}, even if the
environment of the beams, which includes the SQUID, has a higher
temperature.

If the interaction term in Eq. (\ref{eq09}) is small relative to
the other two terms, perturbation theory gives first-order energy
corrections in $\widetilde{V}$ only when $|\omega_3-\omega_4|\ll
\widetilde{V}/\hbar$. Thus, we will henceforth assume the beams
are identical. The Hamiltonian (\ref{eq09}) is quadratic in the
ladder operators and is thus amenable to an exact solution at all
interaction strengths\cite{Bruus04}. This solution is found by
moving to the differential representation and then diagonalizing
the quadratic form of the potential by a canonical transformation
to even and odd coordinates,
\begin{equation}
x_{e,o}=\frac{1}{\sqrt{2}}(x_1\pm x_2),\qquad
p_{e,o}=\frac{1}{\sqrt{2}}(p_1\pm p_2),\label{eq11-2}
\end{equation}
where $x_i$ and $p_i$ are the position and momentum coordinates of
the $i$th beam.

Applying Eq. (\ref{eq11-2}) on the ladder operators, we find
\begin{widetext}
\begin{eqnarray}
a_5=\frac{1}{2\sqrt{2}}\left[(\sqrt{\frac{\omega_5}{\omega_3}}+\sqrt{\frac{\omega_3}{\omega_5}})(a_3+a_4)+(\sqrt{\frac{\omega_5}{\omega_3}}-\sqrt{\frac{\omega_3}{\omega_5}})(a_3^\dagger+a_4^\dagger)\right],\label{eq11-3}\\
a_6=\frac{1}{2\sqrt{2}}\left[(\sqrt{\frac{\omega_6}{\omega_3}}+\sqrt{\frac{\omega_3}{\omega_6}})(a_3-a_4)+(\sqrt{\frac{\omega_6}{\omega_3}}-\sqrt{\frac{\omega_3}{\omega_6}})(a_3^\dagger-a_4^\dagger)\right],\label{eq11-4}
\end{eqnarray}
\end{widetext}
where $a_5^\dagger$ corresponds to creation of an even mode
quantum in which both beams oscillate in phase, and $a_6^\dagger$
to creation of an odd mode quantum, where the beams oscillate in
anti-phase. The even and odd oscillation frequencies are given by
\begin{equation}
\omega_{5,6}=\sqrt{\omega_3^2\pm2\widetilde{V}\omega_3/\hbar}.\label{eq11-5}
\end{equation}

Using this transformation and omitting constant terms, the
Hamiltonian (\ref{eq09}) is reduced to
\begin{equation}
H=\hbar\omega_5 a_5^\dagger a_5+\hbar\omega_6 a_6^\dagger
a_6.\label{eq12}
\end{equation}
We see that the even mode is decoupled from the odd mode in this
Hamiltonian, which is thus separable to an even and an odd part.
The energy spectrum of this Hamiltonian is given by
\begin{equation}
E_{nm}=n\hbar\omega_5 +m\hbar\omega_6 ,\label{eq13}
\end{equation}
while the eigenstates are
\begin{eqnarray}
|nm\rangle=\frac{1}{\sqrt{n!m!}}(a_5^\dagger)^n
(a_6^\dagger)^m|00\rangle,\label{eq14}
\end{eqnarray}
which, upon substitution of Eqs. (\ref{eq11-3}) and
(\ref{eq11-4}), are seen to be highly entangled states of the two
beams.

The quantum state of the system is read by measuring the magnetic
flux threading through the SQUID, which is done by a current
measurement in the standard way\cite{Poole07}. The operator for
this observable is
\begin{equation}
\widehat{\Phi}=-(B+B_1)l_1x_1-(B+B_2)l_2x_2,\label{eq16}
\end{equation}
where the constant term $BA$ was omitted. With the assumption of
$B_1=B_2$, we have
\begin{equation}
\widehat{\Phi}=-\sqrt{2}(B+B_1)l_1\lambda_3(a_5+a_5^\dagger),\label{eq16.5}
\end{equation}
where the zero-point fluctuation, the resonator displacement
uncertainty at the ground state, is defined as
$\lambda_i=\sqrt{\hbar/2m_1\omega_{i+2}}$ with the definition
extended also for $i=3,4$.

We thus see, as expected, that the measurement of the magnetic
flux cannot detect the odd mode, since oscillations in this mode
do not amount to a change in the area of the SQUID loop. We
therefore set to read and write quantum information only in the
$n$ quantum number in the state $|nm\rangle$. Moreover, we note
that for the eigenstates of the Hamiltonian we have $\langle
\widehat{\Phi}\rangle=0$, implying that a better observable would
be the standard deviation $\langle \Delta \widehat{\Phi}\rangle$.
This is indeed true, with the values of this observable on the
eigenstates being
\begin{equation}
\langle \Delta
\widehat{\Phi}\rangle=\sqrt{2}(B+B_1)l_1\lambda_3\sqrt{1+2n},\label{eq17}
\end{equation}
enabling us to measure the value of $n$.

Having established the reading process, we now set to describe how
to write quantum information on this system. It would seem the
best way to excite the system is via a resonant ac current of
frequency $\omega_5$ in the external wires that, according to the
Hamiltonian (\ref{eq12}), will pump the beams to their excited
state. However, such a current will also pump the beams to even
higher excited states, since the energy level difference is fixed
in this system. A better method would be to use constant currents
in the external wires. The addition to the potential
(\ref{eq02-2}) due to such currents, keeping only first-order
terms in $B_1/B$ and $B_2/B$, is
\begin{eqnarray}
H_1=\frac{\pi E_J l_1}{I_c
\beta_L\Phi_0}\{-B_1x_1[4I_c(\overline{\phi}-\phi_e)+I_b\beta_L]\nonumber\\
+B_2x_2[I_b\beta_L-4I_c(\overline{\phi}-\phi_e)]\}\label{eq18}
\end{eqnarray}
where constant terms were omitted and only linear terms in $x_i$
were kept, owing to the quadratic terms being smaller by several
orders of magnitude.

Since reading can be done only for the even mode, and the
Hamiltonian (\ref{eq12}) is separable into odd and even
components, we do not consider the odd part in Eq. (\ref{eq18}),
and by choosing also $B_1=B_2$, the even part of Eq. (\ref{eq18})
is
\begin{eqnarray}
H_{1,e}&=&-4\sqrt{2}\pi\frac{E_J l_1 \lambda_3(\overline{\phi}-\phi_e)}{\beta_L\Phi_0}B_1(a_5+a_5^\dagger)\nonumber\\
   &\equiv&f(B_1)(a_5+a_5^\dagger),\label{eq18.5}
\end{eqnarray}
where constant terms were again omitted. The even part of the
total Hamiltonian $H+H_1$ is therefore
\begin{equation}
H_e=\hbar\omega_5 a_5^\dagger
a_5+f(B_1)(a_5+a_5^\dagger).\label{eq19}
\end{equation}

The idea of writing is then the following: We create a constant
magnetic field $B_1$, which shifts the harmonic potential and then
let the system relax to its new ground state. We then suddenly
revert $B_1$ to zero thereby obtaining an excited state of the
original system. The Hamiltonian in the differential
representation is
\begin{equation}
H_e=-\frac{\hbar^2}{2m_1}\frac{d^2}{d
x^2}+\frac{1}{2}m_1\omega_5^2x^2+\lambda_3^{-1}f(B_1)x,\label{eq20}
\end{equation}
where the mass $m_1$ is the beam mass. Apart from a constant, the
Hamiltonian (\ref{eq20}) is
\begin{eqnarray}
H_e&=&-\frac{\hbar^2}{2m_1}\frac{d^2}{d
x^2}+\frac{1}{2}m_1\omega_5^2(x+\Delta
x)^2,\label{eq21}\\
\Delta x&=&\frac{f(B_1)}{m_1 \lambda_3\omega_5^2}.\label{eq22}
\end{eqnarray}

\begin{figure}[t]
 \begin{center}
 \includegraphics[width=9.00cm]{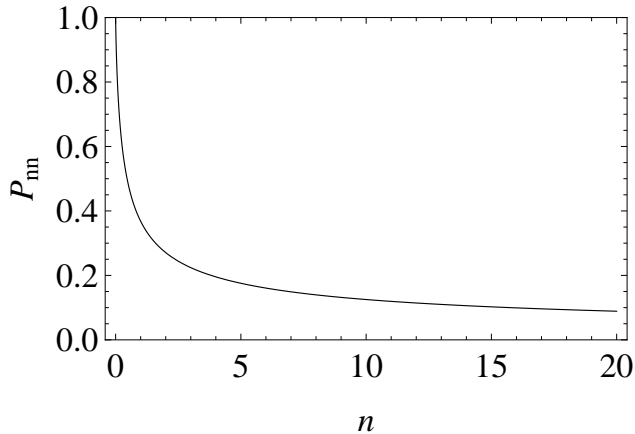}
 \caption{\label{fig02} Graph of the maximum probability of the ground state of the shifted harmonic potential having a $|n\rangle$ component in the unshifted potential. We have $P_{00}=1$ and $P_{11}=e^{-1}$. The values for non-integer $n$ were interpolated by $n!=\Gamma(n+1)$.}
 \end{center}
\end{figure}

According to the scheme described above, we wish to maximize the
probability $P_{nn}=|\langle \psi_0(x+\Delta
x)|\psi_n(x)\rangle|^2$, with $\psi_n(x)$ being the $n$th harmonic
oscillator wavefunction, to obtain the desired state $|n\rangle$
by tuning $B_1$ and with it $\Delta x$. Using standard results of
the quantum harmonic oscillator to write the integral $\langle
\psi_0(x+\Delta x)|\psi_n(x)\rangle$ and then using the generating
function of the Hermite polynomials to find its value for every
$n$, we find the maximum value of $P_{nn}$ is reached when $\Delta
x=2\sqrt{n}\lambda_3$ with the probability then to measure $k$
phonons after $B_1$ is removed given by
\begin{equation}
P_{kn}=(k!)^{-1}e^{-n}n^k,\label{eq23}
\end{equation}
which is a Poisson distribution with mean $n$. The $P_{nn}$
function is plotted in Fig. \ref{fig02}. We see that the probability
drops sharply for small $n$ and evens out for larger values. The
limit at $n\rightarrow\infty$ is 0. Although the writing process
does not create a pure number state $|n\rangle$, the standard
deviation in the number of phonons, by the properties of the
Poisson distribution, is $\sqrt{n}$, which is reasonably low.
%

\section{Squeezed states} \label{section3}
Having shown how to read and write quantum information in this
system, we now wish to demonstrate the possibility of creating
squeezed states. In an effort to mimic the Hamiltonian of a
degenerate parametric amplifier from quantum optics\cite{Walls08},
we set the external wires magnetic fields to oscillate at double
the frequency of the even mode, namely,
\begin{eqnarray}
B_1&=&B_{1,0}e^{2i\langle\omega_5\rangle t}+\mathrm{c.c},\nonumber\\
B_2&=&B_{2,0}e^{2i\langle\omega_5\rangle
t}+\mathrm{c.c},\label{eq24}
\end{eqnarray}
where $\langle \omega_5\rangle$ denotes the time average of
$\omega_5$, and the oscillations of $\omega_5$ about this average
are small since $B_i\ll B$.

The non-interacting Hamiltonian is found from Eq. (\ref{eq12}) to
be
\begin{eqnarray}
H_0=\hbar \langle\omega_5\rangle a_5^\dagger a_5+\hbar
\langle\omega_6\rangle a_6^\dagger a_6.\label{eq25}
\end{eqnarray}
Keeping only terms to first order in $B_i/B$, the addition to the
potential (\ref{eq02-2}) due to the oscillating magnetic field in
the rotating wave approximation (RWA), valid here due to the weak
damping, is
\begin{eqnarray}
\hspace{-2em} H_1=\frac{4 \pi E_J l_1^2 B}{\Phi _0^2 \beta
   _L}\{(B_1+B_2)x_1x_2+B_1 x_1^2+B_2 x_2^2\}.\label{eq26}
\end{eqnarray}

We write this addition in terms of ladder operators, take
$B_{1,0}=B_{2,0}$ to be pure imaginary and eliminate constant
terms to find the interaction picture Hamiltonian in the RWA to be
\begin{eqnarray}
&H_I=\frac{8\pi i\langle\lambda_3\rangle^2 E_J l_1^2 B
|B_{1,0}|}{\Phi_0^2\beta_L}\{a_5^2-(a_5^\dagger)^2\},\label{eq27}
\end{eqnarray}
which is the squeezing Hamiltonian. In writing Eq. (\ref{eq27}) we
neglected the time-dependent term $(\omega_5-\langle
\omega_5\rangle)a_5^\dagger a_5+(\omega_6-\langle
\omega_6\rangle)a_6^\dagger a_6$, since it is negligible relative
to the squeezing term under the assumed experimental conditions.
It is interesting to note that squeezing of the odd mode is not
possible using this scheme, even if $\omega_5$ is replaced with
$\omega_6$ in Eq. (\ref{eq24}). The fundamental reason for this is
Lenz law, which makes the coefficient of $x_1x_2$ in Eq.
(\ref{eq26}) positive and thus excludes terms proportional to
$a_6^2$ in Eq. (\ref{eq27}). The coefficient is positive because
moving both beams in the positive direction costs energy, since
both movements decrease the magnetic flux.

We now consider the effect of squeezing in this system and devise
means to observe it. We assume dissipation is weak, and both beams
are initially in the ground state. Conforming to standard
notation, the squeezing parameter is
\begin{eqnarray}
g=\frac{16\pi \langle\lambda_3\rangle^2 E_J l_1^2 B
|B_{1,0}|t}{\hbar\Phi_0^2\beta_L},\label{eq28}
\end{eqnarray}
which is real, and the squeezing operator reads
\begin{eqnarray}
S(g)=\exp[g(a_5^2-(a_5^\dagger)^2)/2].\label{eq29}
\end{eqnarray}
The time evolution of the $a_5$ operator in the interaction
picture is given by
\begin{eqnarray}
S^\dagger(g)a_5 S(g)=a_5\cosh g-a_5^\dagger \sinh g.\label{eq30}
\end{eqnarray}

In the rotating frame the uncertainty in the positions and momenta
of the beams are
\begin{eqnarray}
\langle\Delta x_i\rangle&=&\lambda_4(1+\tanh (g+\ln(\omega_5/\omega_6)/2))^{-1/2},\label{eq31}\\
\langle\Delta p_i\rangle&=& \frac{\hbar}{2} \lambda_4^{-1}(1-\tanh
(g+\ln(\omega_5/\omega_6)/2))^{-1/2},\label{eq32}
\end{eqnarray}
where Eqs. (\ref{eq05}), (\ref{eq07-1}), (\ref{eq11-3}),
(\ref{eq11-4}) and (\ref{eq30}) were used. We see that we have
limited squeezing to below the standard quantum limit in the
positions and unlimited anti-squeezing in the momenta, as the even
mode is squeezed, while the odd mode is not. The interaction
modifies the squeezing by adding the positive term of
$\ln(\omega_5/\omega_6)/2$ to the squeezing parameter. In
addition, with the product of the uncertainties being
\begin{eqnarray}
\langle \Delta x_i \rangle \langle\Delta
p_i\rangle=\frac{\hbar}{2}\cosh(g+\ln(\omega_5/\omega_6)/2),\label{eq33}
\end{eqnarray}
we see that due to the interaction the minimum uncertainty is no
longer attained before squeezing takes place.

It is interesting to write the wavefunction for the beams in the
differential representation in the presence of squeezing. This
wavefunction can be found by using the relation between the
two-photon coherent states and the squeezed states\cite{Walls08}
to write for the squeezed state $|g\rangle$
\begin{eqnarray}
(\cosh g a_5+\sinh g a_5^\dagger)a_6|g\rangle=0.\label{eq34}
\end{eqnarray}
After moving to the differential representation, Eq. (\ref{eq34})
is solved to find a wavefunction of the form
\begin{equation}
\psi(x_e,x_o)=C_1e^{-\frac{m_1\omega_5}{2\hbar}e^{2g}x_e^2}e^{-\frac{m_1\omega_6}{2\hbar}x_o^2},\label{eq34-2}
\end{equation}
with $C_1$ being a normalizing constant. When we transform this
wavefunction to the beam coordinates via Eq. (\ref{eq11-2}) we
find that the new wavefunction is in a jointly Gaussian form,
namely
\begin{equation}
\psi(x_1,x_2)=C_1\exp(-\frac{\frac{x_1^2}{\sigma_1^2}+\frac{x_2^2}{\sigma_2^2}-\frac{2
r x_1 x_2}{\sigma_1\sigma_2}}{2(1-r^2)}),\label{eq35}
\end{equation}
where $\sigma_1=\sigma_2=\sqrt{2}\langle\Delta x_1\rangle$, with
$\langle\Delta x_1\rangle$ given by Eq. (\ref{eq31}), and the
correlation coefficient given by
\begin{equation}
r=-\tanh (g+\ln(\omega_5/\omega_6)/2).\label{eq36}
\end{equation}

We note that the factor of $\sqrt{2}$ in $\sigma_i$ comes from
$|\psi(x_1,x_2)|^2$, rather than $\psi(x_1,x_2)$, being the
probability distribution. In addition, we see that, as before, the
beam interaction results in an addition to the squeezing
parameter, which gives negative correlation even at $t=0$. The
correlation due to the squeezing is negative, because the
influence of the odd mode, which is not squeezed, increases with
time, producing perfect anti-correlation when the squeezing
parameter goes to infinity.

Lastly, we consider the effect of the squeezing on the measurement
of the magnetic flux in the SQUID. Using Eq. (\ref{eq16.5}), we
find the standard deviation of $\widehat{\Phi}$ in the rotating
frame to be
\begin{eqnarray}
\langle \Delta\widehat{\Phi}\rangle=\sqrt{2}(B+B_1)l_1\lambda_3
e^{-g},\label{eq37}
\end{eqnarray}
which is fully squeezed, while in the lab frame we have
\begin{eqnarray}
\langle
\Delta\widehat{\Phi}\rangle=\sqrt{2}(B+B_1)l_1\lambda_3\nonumber\\
\times\sqrt{\cosh(2g)-\sinh(2g)\cos(2\omega_5 t)},\label{eq38}
\end{eqnarray}
which characteristically oscillates between fully squeezed values
at $t=(\pi/\omega_5)p$, corresponding to Eq. (\ref{eq37}), and
fully anti-squeezed values at $t=(\pi/\omega_5)(p+1/2)$, where $p$
is an integer. We conclude that the squeezing effect is
measurable, and that the squeezing parameter can be found from the
measurements.
\section{Effect of Mechanical Damping} \label{section4}
In reality, the damping of the beam oscillations is weak but
nonzero. With regard to reading and writing quantum information,
this is not a problem, so long as the reading or writing is
performed within a period much shorter than the characteristic
decay time. The squeezed states, however, are measurably degraded even by very weak dissipation, as we show in this
section.

Many models were devised for describing dissipation in quantum
systems\cite{Gardiner04}. We choose here to work with the quantum
master equation. In the interaction picture with the Hamiltonian
(\ref{eq27}), the quantum master equation takes the
form\cite{Chiao08}
\begin{eqnarray}
&\frac{\partial}{\partial
t}\rho(t)=\mathcal{L}_S\rho(t)+\mathcal{L}_{dis}\rho(t),\label{eq39}\\
&\mathcal{L}_S\rho(t)=\half \zeta[a^2-(a^\dagger)^2,\rho(t)],\label{eq40}\\
&\mathcal{L}_{dis}\rho(t)=-\frac{\gamma}{2}(n_{cav}+1)\{[a^\dagger,a\rho(t)]+[\rho(t)a^\dagger,a]\}\nonumber\\
&-\frac{\gamma}{2}n_{cav}\{[a,a^\dagger\rho(t)]+[\rho(t)a,a^\dagger]\},\label{eq41}
\end{eqnarray}
where $\zeta=g/t$ is the squeezing rate and for brevity we write
$a$ instead of $a_5$ and $\omega$ instead of $\omega_5$. In Eqs.
(\ref{eq39}-\ref{eq41}) $\rho(t)$ is the statistical operator for
the system, $\mathcal{L}_S$ and $\mathcal{L}_{dis}$ are the
Liouville operators for the squeezing and dissipation,
respectively, $\gamma=\omega/Q$ is the damping rate of the even
mode, where $Q$ is the beam quality factor, and
$n_{cav}=(e^{\hbar\omega/k_BT}-1)^{-1}$ is the average phonon
occupation in the even mode.

The system can be equivalently described by the Wigner
quasi-probability distribution $W(\alpha,\alpha^*)$ instead of by
the statistical operator $\rho(t)$, where we omit the explicit
time dependence in $W(\alpha,\alpha^*)$ to make the notation
concise. The parameter $\alpha=X_1+iX_2$ is a complex number that
is related to the phase space coordinates via
$\alpha=\frac{1}{2\lambda}x+\frac{i\lambda}{\hbar}p$, where $x$
and $p$ are the even mode position and momentum coordinates,
respectively, and $\lambda=\sqrt{\hbar/2m_1\omega}$ as before. We
convert\cite{Gardiner04,Rabl04} Eq. (\ref{eq39}) to an equation
for the Wigner distribution to find
%
%
%
\begin{eqnarray}
&\frac{\partial W(X_1,X_2)}{\partial
t}=[\zeta(X_1\frac{\partial}{\partial
X_1}-X_2\frac{\partial}{\partial
X_2})+\frac{\gamma}{2}(\frac{\partial}{\partial X_1}
X_1+\frac{\partial}{\partial X_2} X_2)\nonumber\\
&+\frac{1}{4}\gamma (n_{cav}+\half)(\frac{\partial^2}{\partial
X_1^2}+\frac{\partial^2}{\partial X_2^2})]W(X_1,X_2),\label{eq42}
\end{eqnarray}
where we note that the original Wigner function
$\overline{W}(x,p)$ is related to the one used here by
$W(\alpha,\alpha^*)=W(X_1,X_2)=2\hbar\overline{W}(x,p)$.

Equation (\ref{eq42}) is seen to be a special case of the
Fokker-Planck equation with $W(\mathbf{u})$ corresponding to the
probability distribution $P(\mathbf{u};t)$, where
$\mathbf{u}=(X_1,X_2)$. Put in this form, the equation can be
formally written as
\begin{eqnarray}
\hspace{-2em}\frac{\partial W(\mathbf{u})}{\partial
t}=-\nabla\cdot
[\mathbf{F}(\mathbf{u})W(\mathbf{u})]+\frac{D_0}{2}\nabla^2
W(\mathbf{u}),\label{eq43}
\end{eqnarray}
where
$\mathbf{F}=(-(\zeta+\frac{\gamma}{2})X_1,(\zeta-\frac{\gamma}{2})X_2)$
and $D_0=\frac{1}{2}\gamma (n_{cav}+\half)$ are the force and
diffusion constant, respectively. Due to the form of the force in
Eq. (\ref{eq43}), we can use separation of variables to break this
equation into two one-dimensional Fokker-Planck equations with
solutions $W_1(X_1)$ and $W_2(X_2)$, where
$W(X_1,X_2)=W_1(X_1)W_2(X_2)$. These solutions are given by
($i=1,2$)
\begin{eqnarray}
W_i(X_i)&=&\frac{1}{\sqrt{2\pi} \sigma_i(t)} \exp
(-\frac{X_i^2}{2\sigma_i^2(t)}),\label{eq44}\\
\sigma_i(t)&=&\sqrt{(\frac{1}{4}-\frac{D_0}{k_i})e^{-k_it}+\frac{D_0}{k_i}},\label{eq45}\\
k_1&=&2\zeta+\gamma,\label{eq46}\\
k_2&=&\gamma-2\zeta,\label{eq47}
\end{eqnarray}
where $k_i$ are the decay rates.

Equations (\ref{eq44}-\ref{eq47}) indicate that a steady-state
solution always exists for $W_1(X_1)$ and is given by Eq.
(\ref{eq44}) with $\sigma_1(t)=\sqrt{D_0/k_1}$. This finite
distribution width corresponds to a saturation in the squeezing in
contrast with the dissipationless case, when the field quadrature
$X_1$ is squeezed without limit.\cite{Walls08} For $W_2(X_2)$ on
the other hand, we have a steady-state solution only at the strong
damping regime, $\gamma>2\zeta$, and this solution exhibits
$\sigma_2(t)=\sqrt{D_0/k_2}$. When the strong damping condition is
not satisfied, $k_2$ is negative, there is no steady state, and
$X_2$ is anti-squeezed as in the dissipationless
case\cite{Walls08}, but at a slower pace since the leading
behavior in $\Delta X_2$ is $e^{(\zeta-\gamma/2)t}$ instead of
$e^{\zeta t}$ as in the dissipationless case.

The knowledge of the Wigner function in Eq. (\ref{eq44}) enables
us to calculate of system properties via the
relation\cite{Gardiner04}
\begin{eqnarray}
\langle \{a^r(a^\dagger)^s\}_{\mathrm{sym}}\rangle=\int d^2\alpha
\alpha^r(\alpha^*)^s W(\alpha,\alpha^*),\label{eq48}
\end{eqnarray}
where $\{\cdot\}_{\mathrm{sym}}$ indicates the average of all the
permutations of the ladder operators, and $d^2\alpha=dX_1dX_2$.
Working in the rotating frame, the resulting uncertainties in the
positions and momenta of the beams read $(i=1,2)$
\begin{eqnarray}
\langle \Delta
x_i\rangle&=&\sqrt{2}\lambda_3\sqrt{\frac{1}{4}\frac{\omega_5}{\omega_6}+\sigma_1(t)^2},\label{eq49}\\
\langle \Delta
p_i\rangle&=&\frac{1}{\sqrt{2}}\hbar\lambda_3^{-1}\sqrt{\frac{1}{4}\frac{\omega_6}{\omega_5}+\sigma_2(t)^2},\label{eq50}
\end{eqnarray}
which reduce to Eqs. (\ref{eq31}-\ref{eq32}) when $\gamma=0$.

We see that the squeezing in the position coordinates, already
limited to $\sqrt{\omega_3/2\omega_6}$ of the standard quantum
limit, $\lambda_1$, in the dissipationless case of Eq.
(\ref{eq31}), is limited here as well with the same limit, where
we take $n_{cav}= 0$ due to the previous assumption of
$\hbar\omega\gg k_BT$. The momenta uncertainties, in comparison,
are anti-squeezed only in the weak damping regime,
$\gamma<2\zeta$, compared with being always anti-squeezed in Eq.
(\ref{eq32}), when there is no damping. As with the quadrature
field $X_2$, the momenta anti-squeezing in the weak damping regime
has a slower rate relative to the dissipationless case with a
leading behavior of $e^{(\zeta-\gamma/2)t}$ vs. $e^{\zeta t}$ for
the dissipationless case. The product of the position and momentum
uncertainties in Eqs. (\ref{eq49}-\ref{eq50}) gives the lowest
uncertainty at $t=0$ and higher values afterwards.

\begin{figure}[t]
 \begin{center}
\large\flushleft{\mbox{(a)}}\\
\vspace{-2.5em}\includegraphics[width=9.00cm]{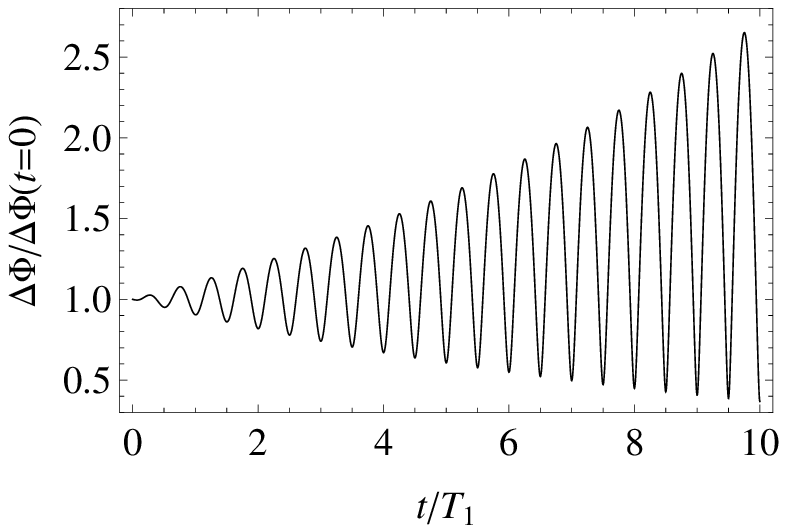}\\
\vspace{-1.1em}\flushleft{\mbox{(b)}}\\
\vspace{-2.5em}\includegraphics[width=9.00cm]{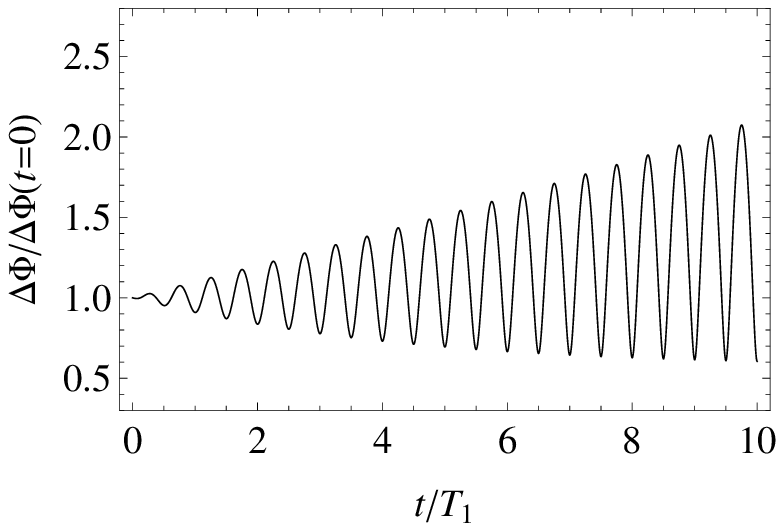}\\
\vspace{-1.1em}\flushleft{\mbox{(c)}}\\
\vspace{-2.5em}\includegraphics[width=9.00cm]{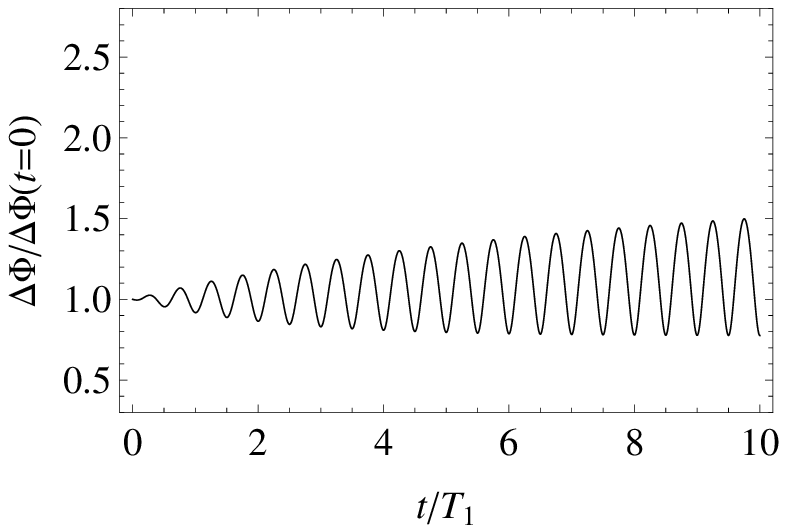}\\
 \caption{\label{fig03} Normalized deviation in the SQUID magnetic flux $\Delta \Phi(t)/\Delta \Phi(t=0)$ as a function of $t/T_1$, where $n_{cav}=0$, $T_1=2\pi/\omega$ and
 $\zeta=0.1/T_1$. (a) No damping ($\gamma=0$): Lower and upper limits are $0$ and $\infty$. (b) Weak damping ($\gamma=\zeta$): Lower and upper limits are $1/\sqrt{3}$ and $\infty$. (c)
 Strong damping ($\gamma=3\zeta$): Lower and upper limits are $\sqrt{3}/2$ and $\sqrt{3}$.}
 \end{center}
\end{figure}
As in Sec. \ref{section3}, we wish to find here the effect of
squeezing on the measurement of the magnetic flux. Using Eq.
(\ref{eq16.5}), we find the standard deviation in the rotating
frame to be
\begin{eqnarray}
\langle \Delta\widehat{\Phi}\rangle=2\sqrt{2}(B+B_1)l_1\lambda_3
\sigma_1(t),\label{eq51}
\end{eqnarray}
which is squeezed, though only to a finite extent unlike the
dissipationless case in Eq. (\ref{eq37}), where it is fully
squeezed. In the lab frame we have
\begin{eqnarray}
\langle
\Delta\widehat{\Phi}\rangle=2\sqrt{2}(B+B_1)l_1\lambda_3\nonumber\\
\times\sqrt{\cos^2(\omega t)\sigma_1^2(t)+\sin^2(\omega
t)\sigma_2^2(t)},\label{eq52}
\end{eqnarray}
which oscillates between squeezed values at $t=(\pi/\omega)p$,
corresponding to Eq. (\ref{eq51}), and squeezed/anti-squeezed
values, depending on the damping regime, at
$t=(\pi/\omega)(p+1/2)$, where $p$ is an integer. In the
dissipationless limit of $\gamma=0$ Eq. (\ref{eq52}) reduces to
Eq. (\ref{eq38}). Eq. (\ref{eq52}) is plotted in Fig. \ref{fig03}
with normalized units for the no-damping, weak-damping and
strong-damping cases. We conclude again that the squeezing effect
is detectible and leads to reduced variation in the measured
magnetic flux in the SQUID.
\section{Conclusions} \label{conclusions}
In this work we have demonstrated that a system composed of two
nanomechanical resonators embedded in a dc SQUID can be used as
two units of quantum memory and that only the even mode in these
two units is readable by the SQUID. We showed how the state of the
beams can be altered, corresponding to writing quantum
information, and proved the amplitude distribution of the number
states in the resulting state is Poisson distributed. We then
proposed a scheme to squeeze the even mode of the resonators and
thus decrease the noise in the SQUID magnetic flux. Taking
dissipation into account, we found a criterion that separates the
weak damping regime, where a steady state exists only in one field
quadrature, from the strong damping one, where both field
quadratures exhibit steady states. We then predicted the form of
the fluctuations in the magnetic flux in the SQUID, by which
squeezing can be observed.

The approximations and assumptions made during our derivations
hold well for reasonable experimental values. For instance, for
two identical 8 MHz resonators of length 25 $\mu$m and quality
factor $Q=2\cdot 10^4$, an external magnetic field of 10 T, beam
temperature of $0.1$ mK, SQUID temperature of 20 mK and other
parameter values similar to the ones in Refs.
\onlinecite{Poot08,Poot10}, we find the energy level differences
in the Hamiltonian (\ref{eq12}) to be much larger than both $k_BT$
and the level widths. Moreover, for the reading process, Eq.
(\ref{eq17}) gives a required SQUID sensitivity of $1.3\cdot
10^{-5}\frac{\Phi_0}{\sqrt{\mathrm{Hz}}}$ for $n\sim 1$ and
sensitivity of $1.3\cdot
10^{-5}\frac{1}{\sqrt{2n}}\frac{\Phi_0}{\sqrt{\mathrm{Hz}}}$ for
$n\gg 1$. A typical SQUID with a flux sensitivity of
$10^{-6}\Phi_0/\sqrt{\mathrm{Hz}}$ satisfies these conditions for
$n<80$.

Regarding the squeezing, a major question is whether substantial
squeezing can be achieved within the decoherence time for the
states. The decoherence time for the resonators here can be made
to be at least 5 $\mu$s\cite{Regal08,Groblacher09,Verhagen12},
while substituting the parameters above in Eq. (\ref{eq28}) gives
a characteristic squeezing time of $\tau_{sq}\sim$ 2 $\mu$s. We
therefore conclude that substantial squeezing is achievable within
the dephasing time.

The experimental realization of this system will be an important
demonstration of macroscopic quantum behavior and squeezing in a
nanomechanical system. In addition it can be used for detecting
the position of the embedded nanomechanical beams with accuracy
higher than the standard quantum limit. Stacking such SQUIDS in
series, with the upper arm of the lower SQUID being also the lower
arm of the upper one, can form a quantum data
bus\cite{Li05,Sillanpaa07}, lead to a multi-mode entangled
state\cite{Nielsen10}, and possibly multi-mode
squeezing\cite{Walls08}. Another application of this system or a
close variant of it is that of a quantum gate\cite{Nielsen10}
acting on the two states by means of currents in the external
wires. A series of such quantum gates can form the basis of a
nanomechanical quantum computer\cite{Cleland04,Ladd10}. We leave
the development of these ideas for future studies.
\section*{Acknowledgments}
This work has been
partially funded by the NSF grant No. DMR-0802830. One of us (MD) is grateful to the Scuola Normale Superiore of Pisa for the hospitality during a visit where part of this
work has been initiated, and to S. Pugnetti and R. Fazio for useful discussions.
\section*{References}
\bibliography{SQUID}
\end{document}